# Mechanical Characterization of Brain Tissue in Simple Shear at Dynamic Strain Rates

Badar Rashid[a], Michel Destrade[b,a], Michael D Gilchrist[a*]

[a]*School of Mechanical and Materials Engineering, University College Dublin, Belfield, Dublin 4, Ireland*

[b]*School of Mathematics, Statistics and Applied Mathematics, National University of Ireland Galway, Galway, Ireland*

*\*Corresponding Author*

Tel: + 353 1 716 1884/1991, + 353 91 49 2344  Fax: + 353 1 283 0534

Email: badar.rashid@ucdconnect.ie (B. Rashid), michel.destrade@nuigalway.ie (M. Destrade), michael.gilchrist@ucd.ie (M.D. Gilchrist)

**Abstract** Mechanical characterization of brain tissue has been investigated extensively over the past fifty years. During severe impact conditions, brain tissue experiences a rapid and complex deformation, which can be seen as a mixture of compression, tension and shear. Moreover, diffuse axonal injury (DAI) occurs in animals and humans when both the strains and strain rates exceed 10% and 10/s, respectively. Knowing the mechanical properties of brain tissue in shear at these strains and strain rates is thus of particular importance, as they can be used in finite element simulations to predict the occurrence of brain injuries under different impact conditions. In this research, an experimental setup was developed to perform simple shear tests on porcine brain tissue at strain rates ≤ 120/s. The maximum measured shear stress at strain rates of 30, 60, 90 and 120/s was 1.15 ± 0.25 kPa, 1.34 ± 0.19 kPa, 2.19 ± 0.225 kPa and 2.52 ± 0.27 kPa, (mean ± SD), respectively at the maximum amount of shear, $K$ = 1. Good agreement of experimental, theoretical (Ogden and Mooney-Rivlin models) and numerical shear stresses was achieved (p = 0.7866 ~ 0.9935). Specimen thickness effects (2.0 – 10.0 mm thick specimens) were also analyzed numerically and we found that there is no significant difference (p = 0.9954) in the shear stress magnitudes, indicating a homogeneous deformation of the specimens during simple shear tests. Stress relaxation tests in simple shear were also conducted at different strain magnitudes (10% - 60% strain) with the average rise time of 14 ms. This allowed us to estimate elastic and viscoelastic parameters ($\mu$ = 4942.0 Pa and Prony parameters: $g_1$ = 0.520, $g_2$ = 0.3057, $\tau_1$ = 0.0264 s, $\tau_2$ = 0.011 s) that can be used in FE software to analyze the hyperviscoelastic behavior of brain tissue. This study will provide new insight into the behavior of brain tissue under dynamic impact conditions, which would assist in developing effective brain injury criteria and adopting efficient countermeasures against TBI.

# 1 Introduction

The human head is the most sensitive region involved in life-threatening injuries due to falls, traffic accidents and sports accidents. Intracranial brain deformations are produced by rapid angular and linear accelerations as a result of blunt impact to the head, leading to traumatic brain injuries (TBIs) which remain a main cause of death and severe disabilities around the world. During severe impact to the head, brain tissue experiences a mixture of compression, tension and shear which may occur in different directions and in different regions of the brain. To gain a better understanding of the mechanisms of TBI, several research groups have developed numerical models which contain detailed geometric descriptions of the anatomical features of the human head, in order to investigate internal dynamic responses to multiple loading conditions (Claessens et al., 1997; Claessens, 1997; Ho and Kleiven, 2009; Horgan and Gilchrist, 2003; Kleiven, 2007; Kleiven and Hardy, 2002; Ruan et al., 1994; Takhounts et al., 2003b; Zhang et al., 2001). However, the biofidelity of these models is highly dependent on the accuracy of the material properties used to model biological tissues; therefore, a systematic investigation the constitutive behavior of brain tissue under impact is essential.

The duration of a typical head impact is of the order of milliseconds. Therefore to model TBI, we need to characterize brain tissue properties over the expected range of loading rate appropriate for potentially injurious circumstances. Diffuse axonal injury (DAI) is characterized by microscopic damage to axons throughout the white matter of the brain, and by focal lesions in the corpus callosum and rostral brainstem and is considered as the most severe form of TBI (Anderson, 2000; Gennarelli et al., 1972; Margulies and Thibault, 1989; Ommaya et al., 1966). DAI in animals and humans has been estimated to occur at macroscopic shear strains of 10% – 50% and strain rates of approximately 10 – 50/s (Margulies et al., 1990; Meaney and Thibault, 1990). Several studies have been conducted to determine the range of strain and strain rates associated with DAI. Bain and Meaney (2000) investigated *in vivo*, tissue-level*,* mechanical thresholds for axonal injury and their predicted threshold strains for injury ranged from 0.13 to 0.34. Similarly, Pfister et al., (2003) developed a uniaxial stretching device to study axonal injury and neural cell death by applying strains within the range of 20%–70% and strain rates within the range of 20 – 90/s to create mild to severe axonal injuries. Bayly et al., (2006) carried out *in vivo* rapid indentation of rat brain to determine strain fields using harmonic phase analysis and tagged MR images. Values of maximum principal strains > 0.20 and strain rates > 40/s were observed in several animals exposed to 2mm impacts of 21 ms duration. Studies conducted by Morrison et al. (2006; 2003; 2000) also suggested that the brain cells are significantly damaged at strains > 0.10 and strain rates > 10/s.

Over the past five decades, several research groups investigated the mechanical properties of brain tissue in order to establish constitutive relationships over a wide range of loading conditions. Mostly dynamic oscillatory shear tests were conducted over a wide frequency range of 0.1 to 10000 Hz (Arbogast et al., 1995; Arbogast and Margulies, 1998; Arbogast et al., 1997; Bilston et al., 1997; Brands et al., 1999; 2000a; 2000b; 2004; Darvish and Crandall, 2001; Fallenstein et al., 1969; Garo et al., 2007; Hirakawa et al., 1981; Hrapko et al., 2008; Nicolle et al., 2004; 2005; Prange and Margulies, 2002; Shen et al., 2006; Shuck and Advani, 1972; Thibault and Margulies, 1998) and various other techniques were used (Atay et al., 2008; LaPlaca et al., 2005; Lippert et al., 2004; Trexler et al., 2011) to determine the shear properties of the brain tissue. Similarly, unconfined compression and tension tests were also performed by various research groups (Cheng and Bilston, 2007; Estes and McElhaney, 1970; Miller and Chinzei, 1997; Miller and Chinzei, 2002; Pervin and Chen, 2009; Prange and Margulies, 2002; Rashid et al., 2012b; Tamura et al., 2008; Tamura et al., 2007; Velardi et al., 2006) to characterize the mechanical behavior of brain tissue at variable strain rates and the reported properties vary from study to study.

Bilston et al. (2001) performed *in vitro* constant strain rate oscillatory simple shear tests on bovine brain tissue (excised from corpus callosum region) using a parallel plate rotational rheometer and achieved strains up to 100% and strain rates of 0.055, 0.234, 0.95/s. Similarly, *in vitro* simple shear tests were performed on human and bovine brain tissue (gray and white matter regions) at a strain rate of 10/s and up to 50% strain by Takhounts et al. (2003a). Hrapko et al. (2006) also performed *in vitro* simple shear experiments on porcine brain tissue (white matter regions) using an oscillatory rotational rheometer with a strain amplitude of 0.01 and frequencies ranging from 0.04 to 16 Hz. The strain rates ranged from 0.01 to 1/s and strains up to 50%. In all these cases, the magnitudes of strain rates are below the axonal injury thresholds except at 10/s strain rate. Only the study conducted by Donnelly and Medige (1997) achieved engineering strain rates of 0, 30, 60 and 90/s with some additional tests performed at 120 and 180/s. Donnelly and Medige performed *in vitro* simple shear tests on cylindrical specimens of human brain tissue; however, these tests were completed within two to five days of postmortem. Therefore, the possibility of significant stiffness changes to the specimen cannot be ruled out due to this long postmortem time duration. Moreover, a two – term power equation ($\sigma = A \varepsilon^B$, where $\sigma$ = shear stress, A = coefficient, B = exponent, $\varepsilon$ = finite shear strain) was used to model the constitutive behavior of tissue. This can be improved upon to model simple shear. In particular, their power law constitutive equation is an odd function of the shear strain only when B is an odd integer, which was not the case in that investigation.

In this study, the mechanical properties of porcine brain tissue have been determined by performing simple shear tests at strain rates of 30, 60, 90 and 120/s. Due to the ready availability of porcine brains, all tests were completed within 8 h of postmortem. A hyperelastic modelling of brain tissue was performed using Fung, Gent, Mooney – Rivlin and one-term Ogden models and the fundamental aspects of simple shear were considered based on recent studies (Destrade et al., 2008; Destrade et al., 2012; Horgan, 1995; Horgan and Murphy, 2011; Horgan and Saccomandi, 2001; Merodio and Ogden, 2005). A hyperviscoelastic analysis was also carried out by performing stress relaxation tests. Numerical simulations were performed in ABAQUS Explicit/6.9 using material parameters from the Mooney – Rivlin and one-term Ogden models in order to analyze the hyperelastic behavior of brain tissue. The experimental challenge with these tests was to attain uniform velocity during simple shearing of the brain tissue. Therefore a High Rate Shear Device (HRSD) was designed to achieve uniform velocity during dynamic tests (Rashid et al., 2012a). This study will provide new insight into the behavior of brain tissue under dynamic impact conditions, which would assist in developing effective brain injury criteria and adopting efficient countermeasures against TBI.

## 2    Materials and Methods

### 2.1    Experimental Setup

A *High Rate Shear Device* (HRSD) was developed in order to perform simple shear tests at dynamic strain rates of 30, 60, 90, 120/s. As shown in Figs 1 (a) and (b), the major components of the apparatus include an *electronic actuator* (700 mm stroke, 1500 mm/s velocity, *LEFB32T-700*, SMC Pneumatics), one ± 5 N *load cell* (rated output: 1.46 mV/V nominal, *GSO series*, Transducer Techniques) and a *Linear Variable Displacement Transducer* (range ± 25 mm, *ACT1000 LVDT*, RDP Electronics). The load cell was calibrated against known masses and a multiplication factor of 13.67 N/V (determined through calibration) was used to convert voltage (V) to force (N). An integrated amplifier (*AD 623 Gain*, G = 100, Analog Devices) with built-in single pole low-pass filters having cut-off frequencies of 10 kHz and 16 kHz was used. The amplified signal was analyzed through a data acquisition system (DAS) with a sampling frequency of 10 kHz.

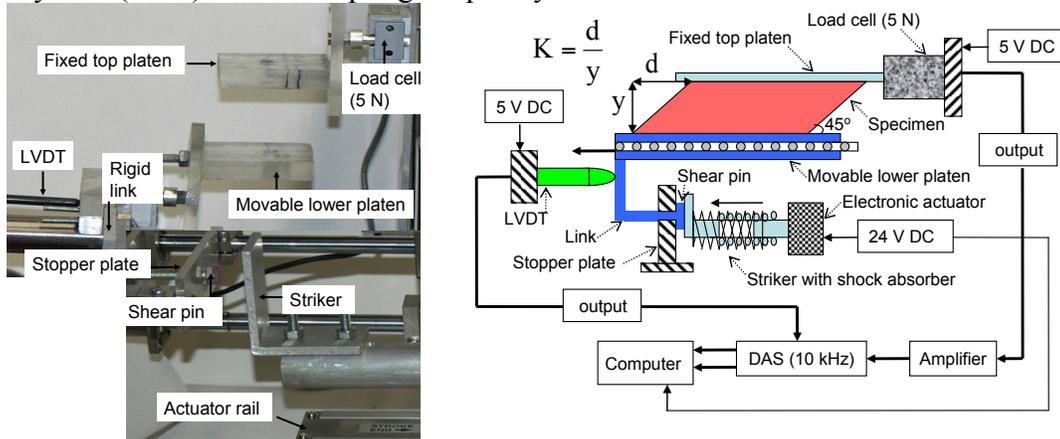

Fig. 1. (a) Major components of high rate shear device (HRSD), (b) Schematic diagram of complete test setup. $K = 1$ for maximum amount of shear

The force (N) and displacement (mm) data against time (s) were recorded for the tissue experiencing the amount of shear, $K = d/y$ ($d$ = displacement of lower platen (mm), $y$ = thickness of specimen (mm)). The *striker* attached to the electronic actuator moved at a particular velocity to strike the *shear pin* which was rigidly attached to the lower platen through a rigid *link* as shown in Fig. 1 (a). During the tests, the top platen remained stationary while the lower platen moved horizontally to produce the required simple shear in the specimen. The two output signals (displacement signal from LVDT and force signal from the load cell), as shown in Fig. 1 (b), were captured simultaneously through the data acquisition system (DAS) at a sampling rate of 10 kHz. The pre-stressed *LVDT probe* was in continuous contact with the *link* to record the displacement during the shear phase of the tests. Two main contributing factors for the non-uniform velocity were the deceleration of the electronic actuator when it approached the end of the stroke and the opposing forces acting against the striking mechanism. Therefore, the striking mechanism was designed and adjusted to ensure that it impacted the tension pin approximately 200 mm before the actuator came to a complete stop. The *striker* impact generated backward thrust, which was fully absorbed by the spring mounted on the actuator guide rod to prevent any damage to the *programmable servo motor*.

## 2.2    Specimen Preparation and Attachment

Ten fresh porcine brains from approximately six month old pigs were collected from a local slaughter house and tested within 8 h postmortem. Each brain was preserved in a physiological saline solution at 4 to 5°C during transportation. All samples were prepared and tested at a nominal room temperature of 23°C. The dura and arachnoid were removed and the cerebral hemispheres were first split into right and left halves by cutting through the corpus callosum and midbrain. As shown in Fig. 2, square specimens composed of mixed white and gray matter were prepared using a square steel cutter with the nominal dimensions (20.0 mm: length x 20 mm: width). The extracted brain specimen was then inserted in a 4.0 mm thick square disk with inner dimension (19.0 mm: length x 19.0 mm: width) as shown in Fig. 2. The excessive brain portion was then removed with a surgical scalpel to maintain an approximate specimen thickness of 4.0±0.1 mm. All specimens were extracted from the cerebral halves while cutting from the medial to lateral direction. Two specimens were extracted from each cerebral hemisphere. The actual thickness of specimens measured before the testing was 4.0±0.2 mm (mean ± SD), respectively. 40 specimens were prepared from 10 brains (4 specimens from each brain).

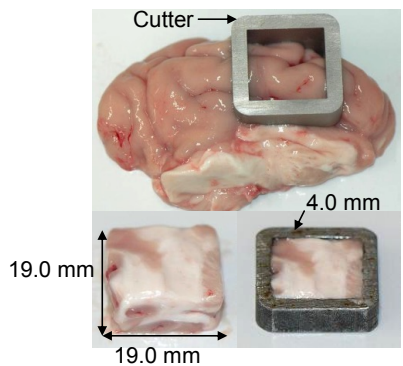

Fig. 2 – Square brain specimen (19.0 ± 0.1 x 19.0 ± 0.1 mm) and 4.0±0.1 mm thick excised from medial to lateral direction. The prepared specimen is placed on a lower platen applied with a thin layer of surgical glue.

The time elapsed between harvesting of the first and the last specimens from each brain was 17 ~ 20 minutes. Due to the softness and tackiness of brain tissue, each specimen was tested only once and no preconditioning was performed (Miller and Chinzei, 1997; Miller and Chinzei, 2002; Tamura et al., 2007; Velardi et al., 2006). Physiological saline solution was applied to specimens frequently during cutting and before the tests in order to prevent dehydration. The specimens were not all excised simultaneously, rather each specimen was tested first and then another specimen was extracted from the cerebral hemisphere. This procedure was important to prevent the tissue from losing some of its stiffness and to prevent dehydration, and thus contributed towards repeatability in the experimentation. Reliable attachment of brain tissue specimens was important in order to achieve high repeatability during simple shear tests. The surfaces of the platens were first covered with a masking tape substrate to which a thin layer of surgical glue (Cyanoacrylate, Low-viscosity Z105880–1EA, Sigma-Aldrich) was applied. The prepared cylindrical specimen of tissue was then placed on the lower platen. The top platen, which was attached to the 5 N load cell, was then lowered slowly so as to just touch the top surface of the specimen. One minute settling time was sufficient to ensure proper adhesion of the specimen to the platens. Before mounting of brain specimens for simple shear tests, calibration of HRSD was essential in order to ensure uniform velocity at each strain rate (30, 60, 90, 120/s). During the calibration process, the actuator was run several times with and without any brain tissue specimen to ensure repeatability of displacement (mm) against time (s).

### 2.3 Stress Relaxation Tests in Simple Shear

A separate set of relaxation experiments were performed on square specimens (19.0 x 19.0 mm: width x length) using a nominal thickness of 4.0 mm. 40 specimens were extracted from 10 brains (4 specimens from each brain). Stress relaxation tests were performed from 10% to 60% engineering strain in order to investigate the viscoelastic behavior of brain tissue. The specimens were shear tested from 30 – 174 mm/s to various strain levels and the data was acquired at a sampling rate of 10 kHz. The average rise time (ramp duration) measured from the stress relaxation experiments was 14 milliseconds (ms) and the shear stress (Pa) vs. time (s) data was recorded up to 150 ms (hold duration). The relaxation data was essentially required for the determination of time-dependent parameters such as $\tau_k$, the characteristic relaxation times, and $g_k$, the relaxation coefficients.

# 3   Phenomenological Constitutive Models

### 3.1   Preliminaries

Let $\underline{\mathbf{F}} = d\underline{\mathbf{x}}/d\underline{\mathbf{X}}$ be the deformation gradient tensor, where $\underline{\mathbf{X}}$ is the position of a material element in the undeformed configuration and $\underline{\mathbf{x}}$ is the corresponding position of the material element in the deformed configuration (Holzapfel, 2008; Ogden, 1997). In the Rectangular Cartesian coordinate system aligned with the edges of the specimen in its undeformed configuration, the simple shear deformation as shown in Fig. 3, can be written as

$x_1 = X_1 + KX_2, x_2 = X_2, x_3 = X_3$ (1)

where $K$ is the amount of shear. Using Eq. (1), the deformation gradient tensor $\underline{\mathbf{F}}$ can be expressed as

$$\underline{\mathbf{F}} = \begin{bmatrix} 1 & K & 0 \\ 0 & 1 & 0 \\ 0 & 0 & 1 \end{bmatrix} \quad (2)$$

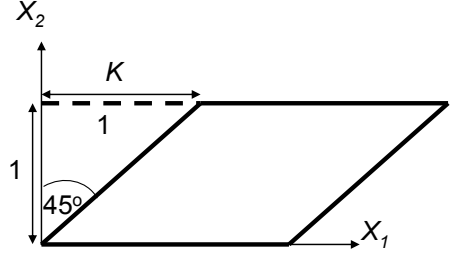

Fig. 3. Schematic of simple shear deformation at amount of shear $K = 1$.

From Eq. (2), the right Cauchy – Green deformation tensor $\underline{\mathbf{C}} = \underline{\mathbf{F}}^T\underline{\mathbf{F}}$ has thus the following components

$$\underline{\mathbf{C}} = \underline{\mathbf{F}}^T\underline{\mathbf{F}} = \begin{bmatrix} 1 & K & 0 \\ K & 1+K^2 & 0 \\ 0 & 0 & 1 \end{bmatrix} \qquad (3)$$

In general, an isotropic hyperelastic incompressible material is characterized by a strain-energy density function $W$ which is a function of two principal strain invariants only: $W = W(I_1, I_2)$, where $I_1$ and $I_2$ are as defined below (Ogden, 1997).

$$I_1 = \mathrm{tr}\,\underline{\mathbf{C}}, \quad I_2 = \frac{1}{2}(I_1^2 - \mathrm{tr}(\underline{\mathbf{C}}^2)) \qquad (4)$$

But in the present case of simple shear deformation,

$$I_1 = I_2 = 3 + K^2 \qquad (5)$$

The shear component of the Cauchy stress tensor is (Ogden, 1997):

$$\sigma_{12} = 2K\left(\frac{\partial W}{\partial I_1} + \frac{\partial W}{\partial I_2}\right) \qquad (6)$$

so that $W = W(3 + K^2, 3 + K^2) \equiv \hat{W}(K)$ say $\qquad (7)$

Using the chain rule, we find

$$\hat{W}' = \frac{\partial(3+K^2)}{\partial K}\frac{\partial W}{\partial I_1} + \frac{\partial(3+K^2)}{\partial K}\frac{\partial W}{\partial I_2} + \frac{\partial(1)}{\partial K}\frac{\partial W}{\partial I_3} = 2K\left(\frac{\partial W}{\partial I_1} + \frac{\partial W}{\partial I_2}\right) = \sigma_{12} \qquad (8)$$

During simple shear tests, the amount of shear $K$, was calculated from the measured displacement of the specimen in the transverse direction and the original thickness of the specimen. The tangential shear stress $\sigma_{12}$ was evaluated as $\sigma_{12} = F/A$, where $F$ is the shear force, measured in Newtons by the load cell, and $A$ is the area of a cross section of the specimen (length: 19.0 mm and width: 19.0 mm). Note that in simple shear, this area remains unchanged. The experimentally measured nominal stress was then compared to the predictions of the hyperelastic models from the relation $\sigma_{12} = \hat{W}'(K)$ (Ogden, 1997), and the material parameters were adjusted to give good curve fitting. Experimental shear stress values and the corresponding amount of shear, $K$ were used for the non-linear least-square fit of the parameters for four common hyperelastic constitutive models, presented in the next sections. The fitting was performed using the ***lsqcurvefit.m*** function in MATLAB, and the quality of fit for each model was assessed based on the coefficient of determination, $R^2$. The fitting of hyperelastic models has been comprehensively covered by Ogden et al. (2004).

## 3.2 Fung Strain Energy Function

The Fung isotropic strain energy (Fung, 1967; Fung et al., 1979) is often used for the modelling of isotropic soft biological tissues. It depends on the first strain invariant only, as follows,

$$W = \frac{\mu}{2b}\left[e^{b(I_1-3)} - 1\right] \tag{9}$$

It yields the following simple shear stress component $\sigma_{12}$ along the $x_1$ – axis,

$$\sigma_{12} = \hat{W}'(K) = \mu K e^{bK^2} \tag{10}$$

Here $\mu > 0$ (infinitesimal shear modulus) and $b > 0$ (stiffening parameter) are the two constant material parameters to be adjusted in the curve-fitting exercise.

## 3.3 Gent Strain Energy Function

The Gent isotropic strain energy (Gent, 1996) is often used to describe rapidly strain-stiffening materials. It also depends on the first strain invariant only in the following manner,

$$W(I_1) = -\frac{\mu}{2}J_m \ln\left(1 - \frac{I_1 - 3}{J_m}\right). \tag{11}$$

It yields the following shear component of the Cauchy stress

$$\sigma_{12} = \hat{W}'(K) = \frac{\mu J_m K}{J_m - K^2}. \tag{12}$$

Here $\mu > 0$ (infinitesimal shear modulus) and $J_m > 0$ are two constant material parameters to be optimized in the fitting exercise.

## 3.4 Mooney - Rivlin Strain Energy Function

Mooney and Rivlin (Holzapfel, 2008; Mooney, 1964; Ogden, 1997) observed that the shear stress response of rubber was linear under simple shear loading conditions. The same concept can be applied to brain tissue also to evaluate the accuracy and predictive capability of this model. It depends on the first and second strain invariants, as follows

$$W(I_1) = C_1(I_1 - 3) + C_2(I_2 - 3) \tag{13}$$

It yields the following shear component of the Cauchy stress tensor,

$$\sigma_{12} = \hat{W}'(K) = 2(C_1 + C_2)K \tag{14}$$

Here $C_1 > 0$ and $C_2 \geq 0$ are two material constants. They are related to the infinitesimal shear modulus as $\mu = 2(C_1 + C_2)$.

## 3.5 Ogden Strain Energy Function

The Ogden model (Ogden, 1972) has been used previously to describe the nonlinear mechanical behavior of brain matter, as well as of other nonlinear soft tissues (Brittany and Margulies, 2006; Lin et al., 2008; Miller and Chinzei, 2002; Prange and Margulies, 2002; Velardi et al., 2006). Soft biological tissue is often modeled well by the Ogden formulation and most of the mechanical test data available for brain tissue in the literature are, in fact, fitted with an Ogden hyperelastic function. The one-term Ogden hyperelastic function is given by

$$W = \frac{2\mu}{\alpha^2}\left(\lambda_1^\alpha + \lambda_2^\alpha + \lambda_3^\alpha - 3\right) \tag{15}$$

where the $\lambda_i$ are the principal stretch ratios (the square roots of the eigenvalues of C, the right Cauchy-green strain tensor).

In simple shear:

$$\lambda_1 = \frac{K}{2} + \sqrt{1+\frac{K^2}{4}},\ \lambda_2 = \lambda_1^{-1} = -\frac{K}{2} + \sqrt{1+\frac{K^2}{4}},\ \lambda_3 = 1 \tag{16}$$

So that

$$\hat{W}(K) = \frac{2\mu}{\alpha^2}\left[\left(\frac{K}{2}+\sqrt{1+\frac{K^2}{4}}\right)^\alpha + \left(-\frac{K}{2}+\sqrt{1+\frac{K^2}{4}}\right)^\alpha - 2\right] \tag{17}$$

and the Cauchy shear stress component $\sigma_{12}$ is

$$\sigma_{12} = \hat{W}'(K) = \frac{\mu}{\alpha}\frac{1}{\sqrt{1+\frac{K^2}{4}}}\left[\left(\frac{K}{2}+\sqrt{1+\frac{K^2}{4}}\right)^\alpha - \left(-\frac{K}{2}+\sqrt{1+\frac{K^2}{4}}\right)^\alpha\right] \tag{18}$$

When $\alpha = 2$, it reduces to $\sigma_{12} = \mu K$ (linear) and recovers the Mooney-Rivlin material with $C_2 = 0$. Here, $\mu > 0$ is the infinitesimal shear modulus, and $\alpha$ is a stiffening parameter.

### 3.6 Viscoelastic Modelling

Biological tissues usually exhibit nonlinear behavior beyond 2 ~ 3% strain. Many nonlinear viscoelastic models have been formulated, but Fung's theory (Fung, 1993) of quasi-linear viscoelasticity (QLV) is probably the most widely used due to its simplicity. To account for the time-dependent mechanical properties of brain tissue, the stress-strain relationship is expressed as a single hereditary integral and a similar approach has been adopted earlier (Elkin et al., 2011; Finan et al., 2012; Miller and Chinzei, 2002). For a Mooney-Rivlin viscoelastic model, we have

$$S(t) = \mu\int_0^t G(t-\tau)\left(\frac{dK}{d\tau}\right)d\tau \tag{19}$$

Here, $S(t)$ is the nominal shear stress component, $\mu = 2(C_1+C_2)$ is the initial shear modulus in the undeformed state, derived from the Mooney-Rivlin model (Eq, 13 and 14), where $C_1 + C_2 > 0$. Note that because the cross-sectional area remains unchanged in simple shear (Ogden, 1997), we have $S(t) = \sigma_{12}(t)$, the Cauchy shear stress component. The relaxation function $G(t)$ is defined in terms of Prony series parameters:

$$G(t) = \left[1 - \sum_{k=1}^n g_k(1-e^{-t/\tau_k})\right] \tag{20}$$

where $\tau_k$ are the characteristic relaxation times, and $g_k$ are the relaxation coefficients. In order to estimate material parameters with a physical meaning, we propose to solve Eq. 19 in two simple steps as discussed in Section 4.3. Similarly, the Ogden-based viscoelastic model can be expressed as follows:

$$S(t) = \frac{\mu}{\alpha} \int_0^t G(t-\tau) \frac{d}{d\tau} \left( \frac{1}{\sqrt{1+\frac{K^2}{4}}} \left[ \left( \frac{K}{2} + \sqrt{1+\frac{K^2}{4}} \right)^\alpha - \left( -\frac{K}{2} + \sqrt{1+\frac{K^2}{4}} \right)^\alpha \right] \right) d\tau \qquad (21)$$

## 4 Results

### 4.1 Experimentation

All simple shear tests were performed on brain specimens containing mixed white and gray matter. The shear tests were performed up to a maximum amount of shear, $K = 1$, i.e., to an angle of 45°. The velocity of the electronic actuator was adjusted to displace the lower platen at the required velocity of 120, 240, 360 and 480 mm/s corresponding to approximate engineering strain rates of 30, 60, 90 and 120/s, respectively. The shear force (N) was sensed by the load cell attached to the top platen as discussed in Section 2.1 and the force – time data obtained at each strain rate was recorded at a sampling rate of 10 kHz. The force (N) was divided by the surface area in the reference configuration to determine shear stress in the tangential direction (along the $x_1$ – axis, as shown in Fig. 4). Similarly, the displacement was divided by the natural thickness of the specimen to determine engineering shear strain. Each specimen was tested once and then discarded because of the highly dissipative nature of brain tissue.

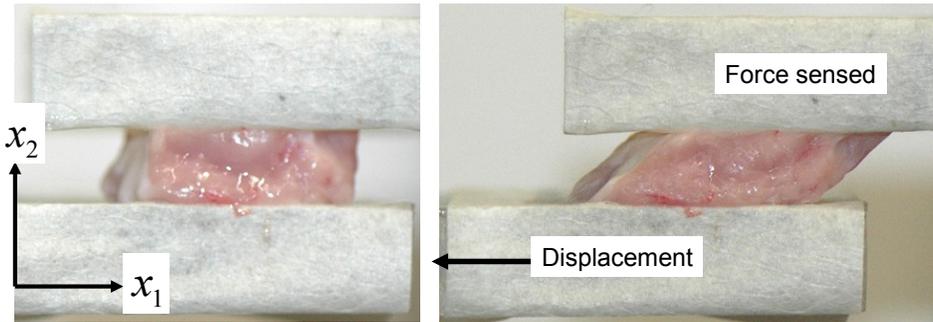

Fig. 4. The brain specimen attached between the platens using thin layer of surgical glue indicating tissue deformation before and after the completion of simple shear test.

Ten tests were performed at each strain rate as shown in Fig. 5, in order to investigate experimental repeatability and the behavior of tissue at a particular loading velocity. The shear force (N) versus time curves increased monotonically at all strain rates.

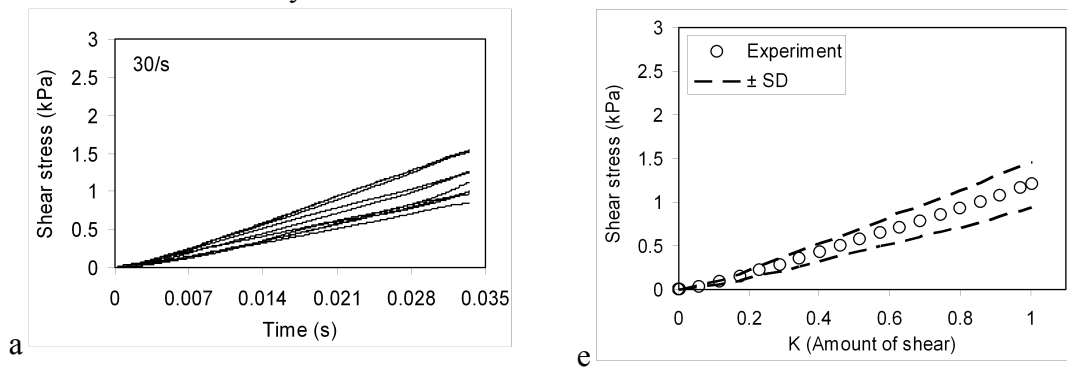

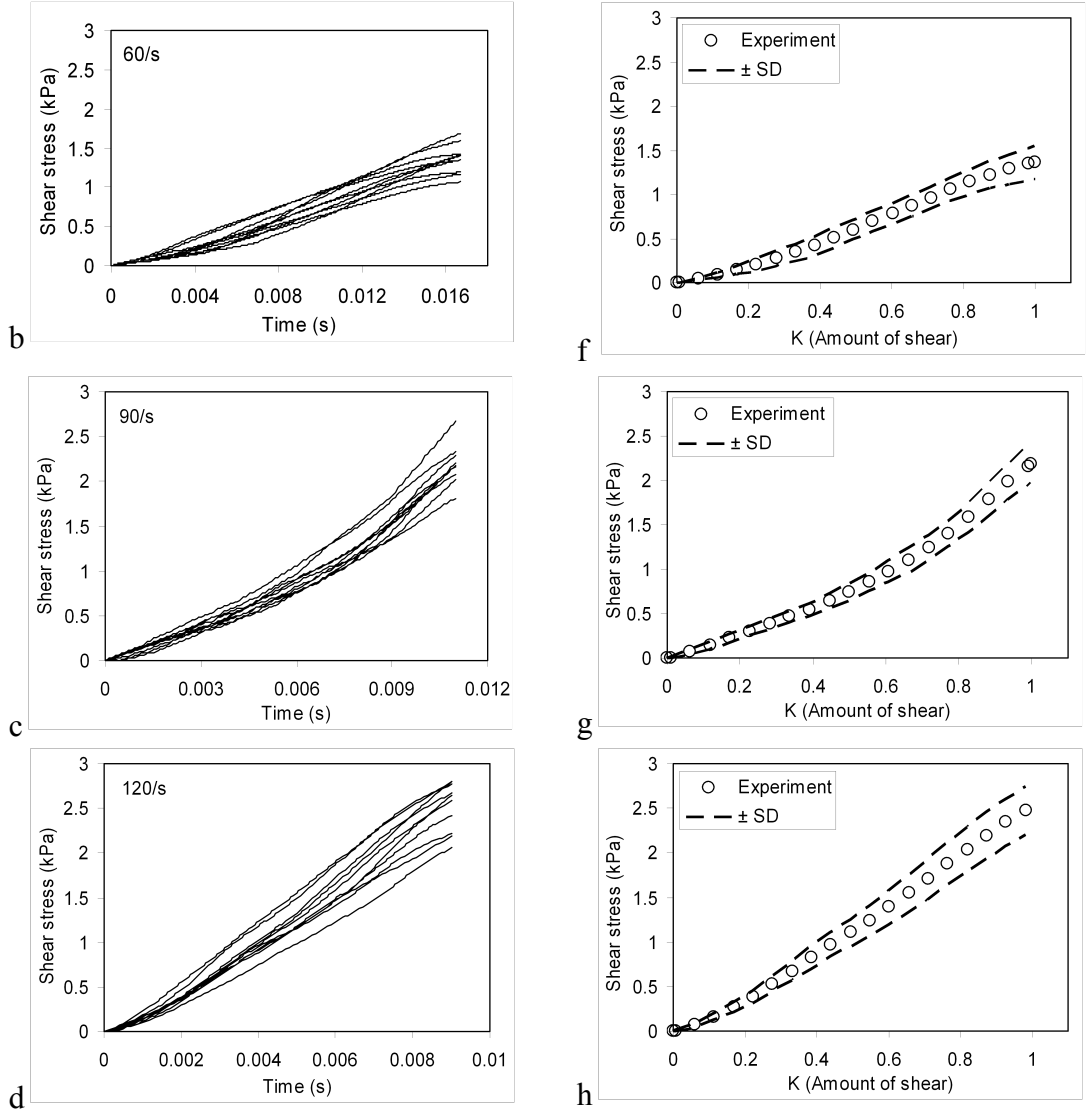

Fig. 5. Experimental shear stress profiles at each strain rate (left) and corresponding average stress and standard deviations (SD) shown on the right side.

During simple shear tests, the achieved strain rates were 30 ± 0.55 /s, 60 ± 1.89 /s, 90 ± 1.78 /s and 120 ± 3.1 /s (mean ± SD) against the required loading velocities of 120, 240, 360 and 480 mm/s, respectively. It was observed that the tissue stiffness increased slightly with the increase in loading velocity, indicating stress – strain rate dependency of brain tissue. Moreover, shear stress profiles are significantly linear at 30, 60 and 120/s strain rates, but this behavior is not quite observed in the case of the 90/s strain rate. The maximum shear stress at strain rates of 30, 60, 90 and 120/s was 1.15 ± 0.25 kPa, 1.34 ± 0.19 kPa, 2.19 ± 0.225 kPa and 2.52 ± 0.27 kPa (mean ± SD), respectively as shown in Fig. 5.

## 4.2  Fitting of Constitutive Models

The average shear stress (Pa) – amount of shear, $K$ (engineering shear strain) curves at each loading rate as shown in Fig. 5, were fitted to the hyperelastic isotropic constitutive models (Fung, Gent, Mooney – Rivlin and Ogden models) discussed in Section 3. Fitting of each constitutive model to experimental data is shown in Fig. 6. Good fitting is achieved for the Fung and Ogden models (coefficient of determination: $0.9883 < R^2 \leq 0.9997$); however, the Mooney – Rivlin and Gent models could not provide as good a fitting ($R^2 < 0.9899$) to shear data, particularly at a strain rate of 90/s due to the nonlinear behavior of the shear stress curve. Material parameters of the four constitutive models ($\mu, \alpha, b, J_m, C_1, C_2$) were derived after fitting to average and standard deviation ($\pm$ SD) shear stress curves and these are summarized in Table 1. Mooney – Rivlin is a linear shear response model ($\sigma_{12} = 2(C_1 + C_2)K$) and is best suited to fit linear experimental shear data as observed at strain rates of 30, 60 and 120/s; however, the Ogden model is suitable for both linear and nonlinear experimental data as shown in Fig.6. If we consider for instance the Ogden model, we see that the initial shear modulus $\mu$ increases to 9.3%, 11.6% and 63.6% with the increase in strain rates from 30 to 60/s, 60 to 90/s and 90 to 120/s, respectively (Table 1). A similar increase in $\mu$ is also observed in the case of the Fung, Gent and Mooney-Rivlin models. The significant increase in $\mu$ with increasing strain rate clearly indicates that a non-linear viscoelastic model is required here. In case of linear viscoelasticity, the behavior is predicted by an increase in the infinitesimal shear modulus, which is different from $\mu$ obtained through Mooney-Rivlin model. Note that the curve fitting exercise for the Mooney-Rivlin model gives only access to $(C_1 + C_2)$, and not to $C_1$ and $C_2$ independently.

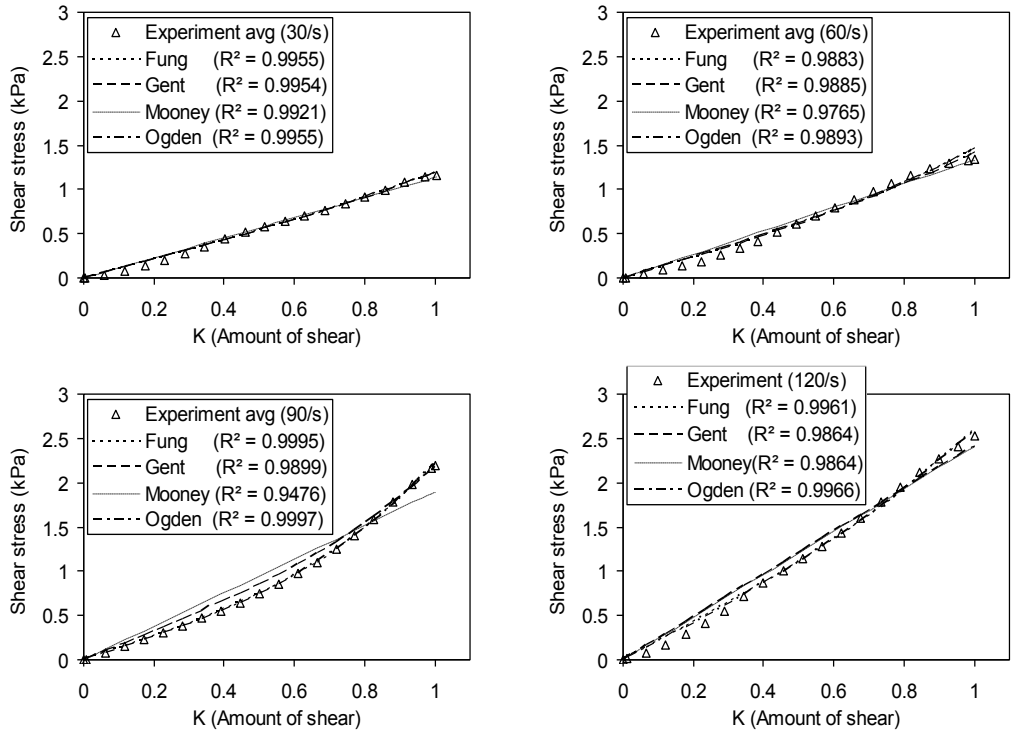

Fig. 6. Fitting of strain energy functions to average experimental shear data at variable strain rates

## 4.3 Estimation of Viscoelastic Parameters

During the stress relaxation tests, the displacement against time is directly recorded through a linear variable displacement transducer (LVDT) and the force (N) is measured directly through the load cell. The relaxation tests were performed at a high loading velocity in order to achieve a minimum rise time, $t$ approximately 14 ms as shown in Fig. 7 (a) and (b). However, it is not practically possible to achieve an ideal step response (time, $t = 0$). Therefore, **back - extrapolation** using *interp1* was performed in order to eliminate any error introduced by the ramp relaxation tests (Funk et al., 2000). Moreover, Laksari et al., (2012) followed a similar approach and determined the instantaneous elastic stress response for the ideal step using a direct numerical integration scheme. However, in this study, extrapolated data was used to generate isochrones (shear stress values at different strain magnitudes but at the same time, $t = 0$) as shown in Fig. 7 (c). Curve fitting of the Mooney – Rivlin model (Eq. 14) was performed using average isochrones to derive $C_1+C_2$, thus directly giving the initial shear modulus $\mu = 2(C_1+C_2)$ which is independent of time as shown in Fig 7 (d). Thereafter, Eq. 19 was convenient to implement in Matlab (Mathworks) by using the *gradient* and *conv* functions. The *gradient* function was used in order to determine the velocity vector $\left(\dfrac{dK}{d\tau}\right)$ from the experimentally measured displacement, $K$ and time, $\tau$. Also a *conv* function was used to convolve the relaxation function (Eq. 20) with the velocity vector, $\left(\dfrac{dK}{d\tau}\right)$. The coefficients of the relaxation function were optimized using *nlinfit* and *lsqcurvefit* to minimize error between the experimental stress data and Eq. (20). The sum of the Mooney-Rivlin constants is $C_1+C_2 = 2471.1$ Pa and the corresponding shear modulus is thus $\mu = 4942.0$ Pa. Similarly, we estimated the Prony parameters ($g_1= 0.520$, $g_2= 0.3057$, $\tau_1 = 0.0264$ s, $\tau_2 = 0.011$ s) from the two-term relaxation function (coefficient of determination: $0.9891 < R^2 \leq 0.9934$) using the Matlab functions discussed above. These material parameters can then be used in FE software such as ABAQUS in order to analyze hyperviscoelastic behavior of brain tissue.

Table 1. Material parameters derived after fitting of models, $\mu$ (Pa), (mean with 95% confidence bound).

| | Fung model | | | Gent model | | |
|---|---|---|---|---|---|---|
| Strain rate (1/s) | $\mu$ | $b$ | $R^2$ | $\mu$ | $J_m$ | $R^2$ |
| 30 | 1047 ± 258 | 0.122 ± 0.022 | 0.9955 | 1050 ± 257 | 9.0 ± 1.8 | 0.9954 |
| 60 | 1157 ± 210 | 0.229 ± 0.04 | 0.9883 | 1197 ± 228 | 6.7 ± 2.2 | 0.9885 |
| 90 | 1322 ± 156 | 0.520 ± 0.03 | 0.9995 | 1594 ± 393 | 3.7 ± 1.23 | 0.9899 |
| 120 | 2104 ± 347 | 0.211 ± 0.067 | 0.9961 | 2414 ± 630 | 5.6 ± 1.3 | 0.9864 |
| | Mooney – Rivlin model | | | Ogden model | | |
| | $C_1+C_2$ | | $R^2$ | $\mu$ | $\alpha$ | $R^2$ |
| 30 | 567.5 ± 207 | | 0.9921 | 1038 ± 258 | 2.766 ± 0.21 | 0.9957 |
| 60 | 665.5 ± 214 | | 0.9765 | 1135 ± 215 | 3.338 ± 0.24 | 0.9893 |
| 90 | 947.7 ± 188 | | 0.9476 | 1267 ± 182 | 4.576 ± 0.18 | 0.9997 |
| 120 | 1166.6 ± 160 | | 0.9864 | 2073 ± 351 | 3.231 ± 0.32 | 0.9966 |

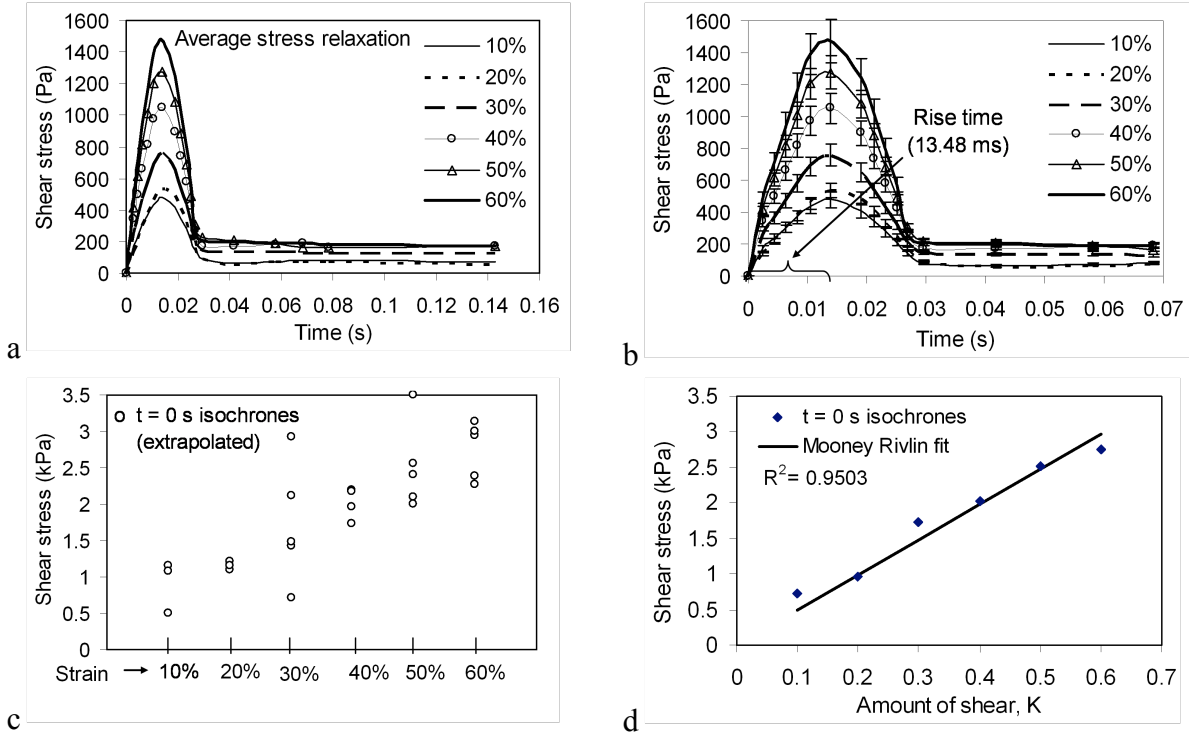

Fig. 7. Stress relaxation experiments in simple shear at different strain magnitudes, with average rise time of 13.48 ms. (a) relaxation data upto 140 ms during hold period (b) average rise time (13.48 ms) at different strains (c) isochronous data after extrapolation (d) fitting of Mooney Rivlin model to average isochrones.

## 5    Finite Element Analysis

### 5.1   Numerical and Experimental Results

A brain tissue specimen geometry; 19.0 mm x 19.0 mm x 4.0 mm (length x width x thickness) was developed using finite element software ABAQUS 6.9/ Explicit for numerical simulations. 2166 x C3D8R elements (Continuum, three – dimensional, 8 – node linear brick, reduced integration) with default hourglass control being used for the brain part. The mass density 1040 kg/m$^3$ and material parameters listed in Table. 1 for the Mooney – Rivlin and Ogden strain energy functions were used for numerical simulations. The top surface of the specimen was constrained in all directions whereas the lower surface was allowed to be displaced only in the lateral direction ($x_1$ - axis) in order to achieve the maximum amount of shear, $K = 1$ for all simulations. Mesh convergence analysis was also carried out by varying mesh density before validating the results. The mesh was considered convergent when there was a negligible change in the numerical solution (0.6%) with further mesh refinement and the average simulation time was approximately 50s.

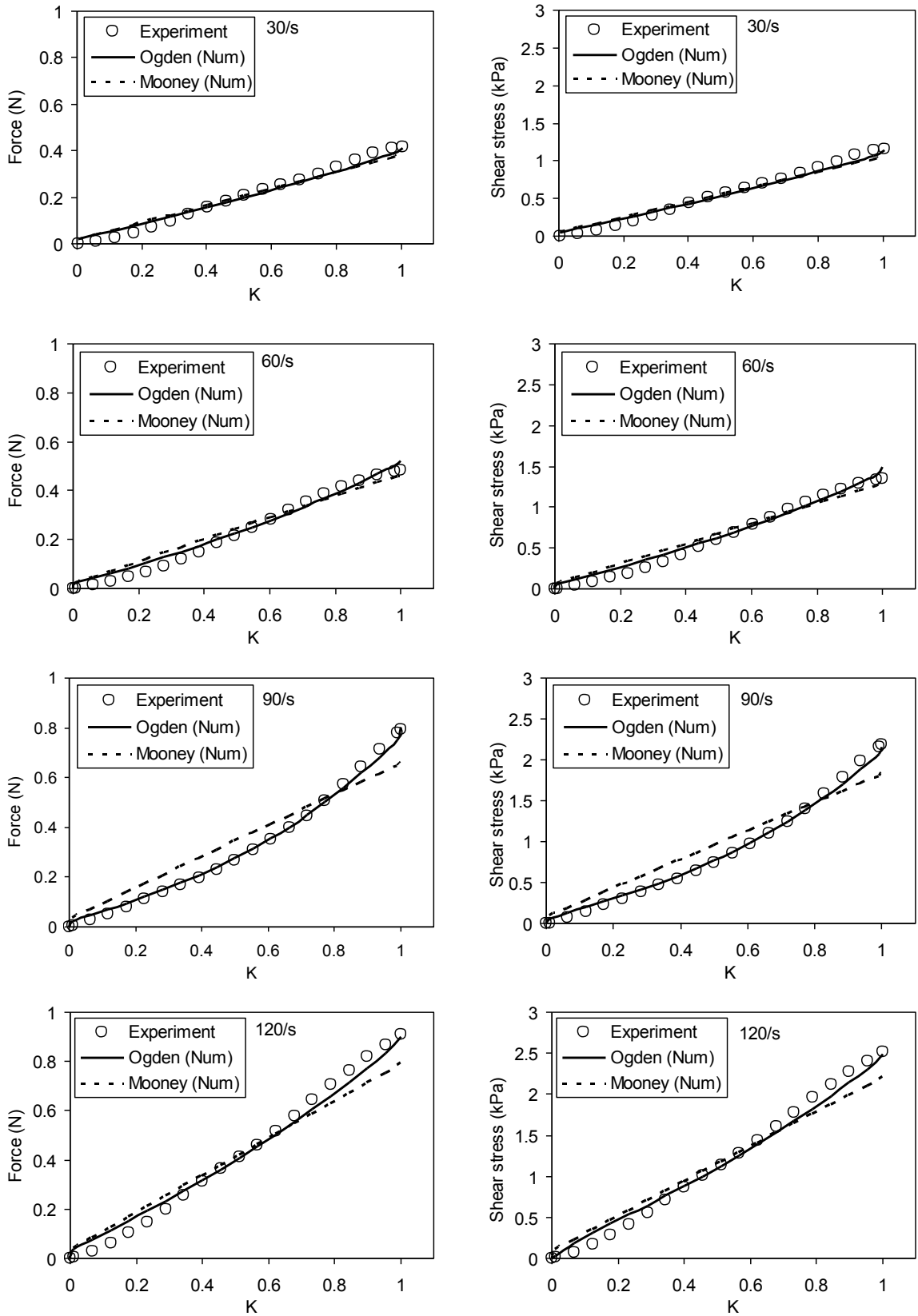

Fig. 8. Comparison of shear forces (N) and shear stresses (kPa) at different strain rates

Simulations were performed in order to determine the force (N) on the top surface of the specimen (along the $x_1$- axis or tangential direction) and were compared with the experimental force (N) measured directly during simple shear tests. A similar procedure was also adopted to compare shear stresses (kPa) as shown in Fig. 8. Based on the statistical analysis using a one-way ANOVA test, a good agreement was achieved between the experimental and numerical results (Ogden and Mooney-Rivlin models) as shown in Table 2.

| Table 2. Statistical comparison of experimental and numerical forces (N) and stresses (kPa) using one-way ANOVA test based on Fig. 7 ||||
|---|---|---|---|---|
| Strain rate (1/s) | Experimental force (N) || Experimental stress (kPa) ||
| | Ogden (numerical) | Mooney (numerical) | Ogden (numerical) | Mooney (numerical) |
| 30/s | p = 0.9793 | p = 0.9864 | p = 0.9791 | p = 0.9803 |
| 60/s | p = 0.8268 | p = 0.8169 | p = 0.8201 | p = 0.7974 |
| 90/s | p = 0.9752 | p = 0.7928 | p = 0.9935 | p = 0.7866 |
| 120/s | p = 0.9950 | p = 0.9338 | p = 0.9830 | p = 0.9553 |

## 5.2 Homogeneous Deformation

Shear stress contours provided by the numerical simulations were also examined. The comparison was carried out for the material parameters at different strain rates using the one-term Ogden model as shown in Fig. 9. The shear stress concentration is conspicuous at the two opposite corners (or edges) on the diagonal with maximum stretched length as depicted in Fig. 9 (a). However, homogeneous stress behavior is observed over the larger volume of the specimen in all the cases. A similar procedure was adopted to compare force (N) contours at each strain rate. The negative force magnitudes are observed on each node at the lower sliding surface of the specimen whereas positive reaction forces are noticed on the top surface of the specimen, which was expected under simple shear deformations (so called Poynting effect, see Ogden (1997)) . The homogeneous force pattern is achieved at each strain rate.

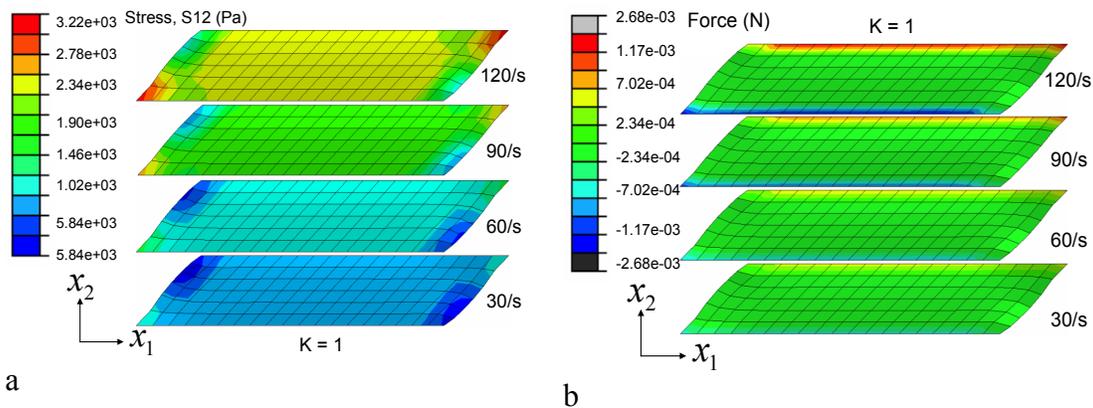

Fig. 9. Simple shear deformation using Ogden parameters (a) shear stress (Pa) contours (b) force (N) contours

## 5.3 Specimen Thickness Effects in Simple Shear

Numerical simulations were also performed at variable specimen thicknesses (2.0, 3.0, 4.0, 7.0 and 10.0 mm) in order to analyze thickness effects. Simulations were performed using the one-term Ogden parameters obtained from the experimental data at a strain rate of 90/s as shown in Fig. 10; however similar behavior can also be observed at other strain rates. Based on a statistical analysis, using a one-way ANOVA test, it is interesting to note that there is no significant difference (p = 0.9954) in the shear stress magnitudes between specimens of different thickness (2.0 – 10.0 mm). The consistency in shear stress magnitudes, as indicated in Fig. 10, clearly indicates homogeneous deformation of the specimen and the results are independent of specimen thickness.

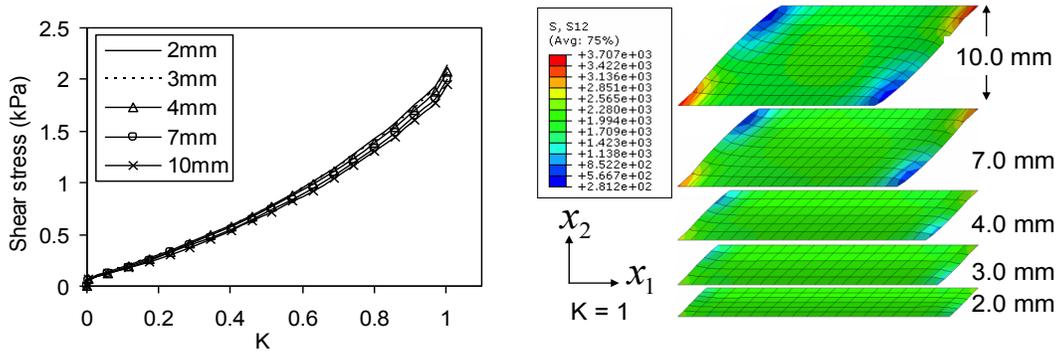

Fig. 10. Consistency in shear stress profiles at variable sample thickness using Ogden material parameters obtained at a strain rate of 90/s

Therefore, the results of the simple shear test protocol can be considered to be much more reliable as compared to the compression and tension test protocols adopted by various research groups (Cheng and Bilston, 2007; Estes and McElhaney, 1970; Miller and Chinzei, 1997; Miller and Chinzei, 2002; Pervin and Chen, 2009; Prange and Margulies, 2002; Rashid et al., 2012b; Tamura et al., 2008; Tamura et al., 2007; Velardi et al., 2006).

# 6 Discussion

Simple shear tests were successfully performed on porcine brain tissue at variable strain rates (30, 60, 90 and 120/s) on a custom designed HRSD, within 8 h postmortem. These strain rates entirely cover the range associated with DAI (Bain and Meaney, 2000; Bayly et al., 2006; Margulies et al., 1990; Meaney and Thibault, 1990; Morrison et al., 2006; 2003; 2000; Pfister et al., 2003). However, in order to obtain one single set of material parameters for the brain tissue, stress relaxation tests were also performed at various strain magnitudes as discussed in Section 4.3. The proposed elastic and viscoelastic parameters ($\mu$ = 4942.0 Pa and Prony parameters: $g_1$ = 0.520, $g_2$ = 0.3057, $\tau_1$ = 0.0264 s, $\tau_2$ = 0.011 s) can be used directly in FE software to analyze the hyperviscoelastic behavior of brain tissue.

Special attention was paid to maintain uniform velocity and a constant strain during the stress relaxation tests at each loading rate by calibrating the HRSD before the tests. Figure 11 shows the start and end of a typical simple shear test indicated between points A and B, respectively. The DAS as discussed in Section 2.1 was able to capture force and displacement signals directly.

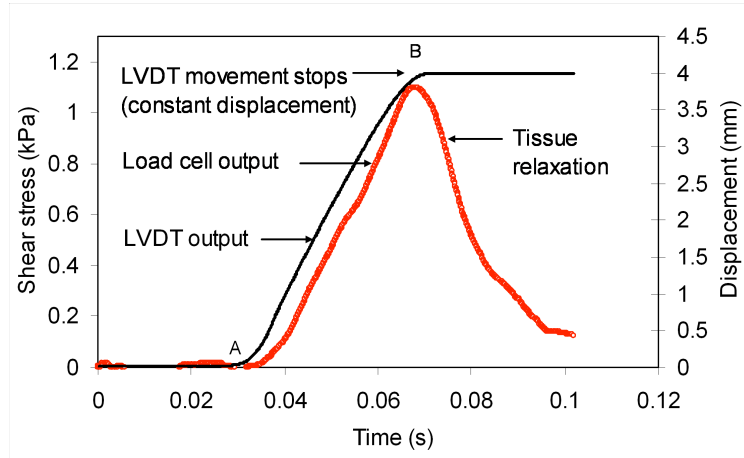

Fig. 11. A typical output from the data acquisition system
(DAS) indicating force (N) and displacement (mm) signals.

During the calibration process, the measured displacement signal was checked precisely against the actual displacement of the shear pin. At this preparatory stage, the velocity of the actuator was adjusted with precision to attain a required strain rate. A linear displacement – time profile between points A and B (~ 4.0 mm displacement) preclude the possibility of any stoppages or irregular movements due to frictional effects between the reciprocating components (see Fig. 11). The movement of the LVDT stops at point B, however the DAS continuously measures the displacement and force signals. The displacement signal measured at point B and onward indicates no relative displacements between the top and bottom platens (acquiring constant strain) during simple shear tests. The brain tissue in simple shear under constant strain conditions starts relaxing with the increase in time (stress relaxation) as shown in Fig. 11. However, separate relaxation tests were performed at different strain magnitudes (10% - 60%).

It is noticed that the response in simple shear is almost linear for most of the strain rates, which somewhat contradicts previous work by e.g. Franceschini et al. (2006). The constants of the Ogden model used by Franceschini et al. lead to a material that is shear-stiffening, whereas the Mooney-Rivlin gives a linear stress - strain curve. But their modelling was based on uniaxial compression / tension of human brain samples with glued ends, for which homogeneous deformations are not possible. However, we have achieved homogeneity in a very satisfying way by performing simple shear tests on porcine brain tissue.

Good agreement was achieved between the theoretical, numerical and experimental results, particularly in the case of the Ogden model, which is deemed suitable for both the linear and nonlinear experimental shear data. However, the Mooney – Rivlin model was good for the linear experimental shear data only. Homogeneous deformation was achieved during simple shear tests and the magnitude of shear stress was proved to be independent of specimen thickness. The maximum shear stress (at $K = 1$) obtained at strain rates of 30, 60, 90 and 120/s was $1.15 \pm 0.25$ kPa, $1.34 \pm 0.19$ kPa, $2.19 \pm 0.225$ kPa, $2.52 \pm 0.27$ kPa, (mean $\pm$ SD), respectively. Donnelly and Medige (1997) also performed in vitro simple shear tests on human brain tissue at the same strain rate range (30 – 90/s); however, the magnitudes of their stresses were 3 to 4 times higher as compared to this study, as shown in Fig. 12.

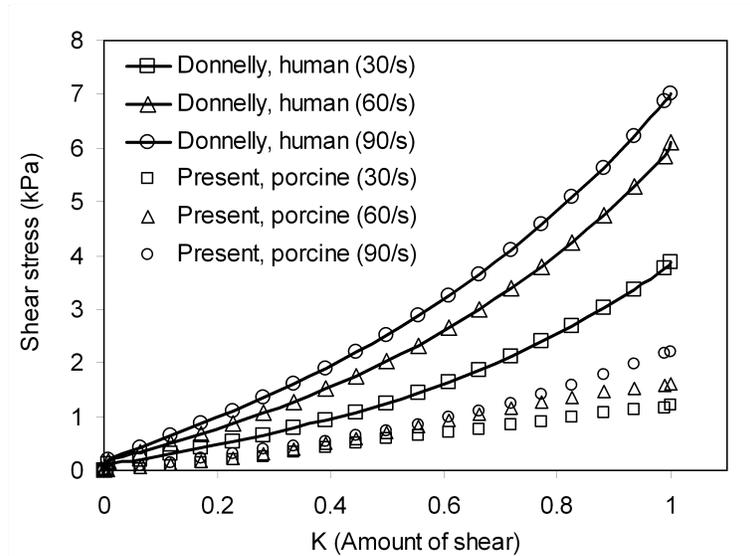

Fig. 12. Comparison of simple shear stress (kPa) profiles with the Donnelly and Medige (1997).

Donnelly completed shear tests within two to five days of postmortem, whereas in this study all tests were completed within 8 h of postmortem. Gefen and Margulies (2004) carried out a comparison between *in vivo* and *in vitro* mechanical behavior of brain tissue and found that the postmortem time for testing was the dominant cause for the large variation in results, whereas pressurized vasculature (during in vivo tests), loss of perfusion pressure (during in vitro tests) and inter-species variability had very little effect on the experimental results. Therefore, the possibility of significantly stiffer specimens cannot be ruled out due to the high postmortem time duration observed in the study of Donnelly and Medige (1997). Another reason for their higher stresses may be due to different species (human and porcine brain tissue). These differences were also identified by Prange et al., (2000) who demonstrated that human brain tissue stiffness was 1.3 times higher than that of porcine brain. However, Nicolle et al., (2004) observed no significant difference between the mechanical properties of human and porcine brain matter. Moreover, the large variations in the experimental data at each strain rate were also observed in the study conducted by Donnelly and Medige (1997), which may be due to large variations in the specimen thicknesses (5.3 ~ 26.4 mm).

Based on numerical simulations, it is observed that the shear stresses are independent of specimen thickness, which shows homogeneous deformation of the brain tissue specimen up to $K = 1$. Therefore, derived material parameters using simple shear test protocol are more reliable than compression and tension test protocols adopted earlier (Cheng and Bilston, 2007; Estes and McElhaney, 1970; Miller and Chinzei, 1997; Miller and Chinzei, 2002; Pervin and Chen, 2009; Prange and Margulies, 2002; Rashid et al., 2012b; Tamura et al., 2008; Tamura et al., 2007; Velardi et al., 2006).

A limitation of this study is that the estimation of material parameters from the strain energy functions is based on average mechanical properties (mixed white and gray matter) of the brain tissue; however, these results are still useful in modelling the approximate behavior of brain tissue. For instance, the average mechanical properties were also determined by Miller and Chinzei (1997; 2002). In previous studies, it was observed that the anatomical origin or location as well as the direction of excision of samples (superior – inferior and medial – lateral direction) had no significant effect on the results (Tamura et al., 2007) and similar observations were also reported by Donnelly and Medige (1997). Therefore inter-regional variations were not investigated in the present research; however inter-specimen variations are clearly evident from the experimental data at each strain rate as shown in Fig.5.

# 7    Conclusions

The following results can be concluded from this study:

1 – Good agreement was achieved between the theoretical, numerical and experimental results, particularly in the case of the Ogden model (p = 0.8201 – 0.9830) which was suitable for representing both linear and nonlinear experimental shear data. However, the Mooney – Rivlin model was good for the linear experimental shear data only (p = 0.7866 – 0.9803).

2 – An approach adopted for the estimation of viscoelastic parameters can be adopted for the Ogden model also. The derived elastic and viscoelastic parameters ($\mu$ = 4942.0 Pa and Prony parameters: $g_1$ = 0.520, $g_2$ = 0.3057, $\tau_1$ = 0.0264 s, $\tau_2$ = 0.011 s) can be used directly in FE software to analyze the hyperviscoelastic behavior of brain tissue.

3 – The high rate shear experimental setup developed for simple shear tests of porcine brain tissue can be used with confidence at dynamic strain rates (30 – 120/s).

**Acknowledgements**    The authors thank Dr. John D. Finan of Columbia University for his valuable input regarding implementation of the hyperviscoelastic model. This work was supported for the first author by a Postgraduate Research Scholarship awarded in 2009 by the Irish Research Council for Science, Engineering and Technology (IRCSET), Ireland.